# Watching Domains Grow: In-situ studies of polarization switching by combined Scanning Probe and Scanning Transmission Electron Microscopy

*Hyejung Chang[1], Sergei V. Kalinin[1], Seungyeul Yang[2], Pu. Yu[2], Saswata Bhattacharya[3], Ping P. Wu[3], Nina Balke[1], Stepehn Jesse[1], Long Q. Chen[3], Ramamoorthy Ramesh[2], Stephen J. Pennycook[1], and Albina Y. Borisevich[1,1]*

[1] Materials Science and Technology Division, Oak Ridge National Laboratory, Oak Ridge, TN 37831, USA

[2] Department of Materials Science and Engineering and Department of Physics, University of California, Berkeley, CA 94720, USA

[3] Department of Materials Science and Engineering, Penn State University, University Park, PA, 16802, USA

---

[1] Corresponding author; Albina Y. Borisevich

albinab@ornl.gov, phone: 865-576-4060, fax: 865-574-6098

Ferroelectric domain nucleation and growth in multiferroic $BiFeO_3$ films is observed directly by applying a local electric field with a conductive tip inside a scanning transmission electron microscope. The nucleation and growth of a ferroelastic domain and its interaction with pre-existing 71° domain walls are observed and compared with the results of phase-field modeling. In particular, a preferential nucleation site and direction-dependent pinning of domain walls is observed due to slow kinetics of metastable switching in the sample without a bottom electrode. These *in-situ* spatially-resolved observations of a first-order bias-induced phase transition reveal the mesoscopic mechanisms underpinning functionality of a wide range of multiferroic materials.

I. Introduction

The switchable polarization in ferroelectric materials enables multiple device applications such as non-volatile random access memories[1] and tunneling barriers.[1,2] Understanding the fundamental physics of ferroelectric domain stability and dynamics in the presence of electric fields is of crucial importance for these applications. Domain switching kinetics in macroscopic ferroelectrics have been extensively studied using classical charge-based measurements, and can generally be described using statistical Kolmogorov-Avrami-Ishibashi (KAI)-type models.[3,4,5] The effects of applied electric field[6,7] and temperature[8] have been explored in detail. However, the concurrent trends for electronic device miniaturization and growth of low-defect density epitaxial films necessitate the understanding of the polarization switching phenomena on the level of a single structural defect, i.e. the development of deterministic predictive models as opposed to statistical description.

Recently, piezoresponse force microscopy (PFM) has emerged as a powerful tool for probing local bias-induced phase transitions in ferroelectrics. In tip-electrode PFM, a local bias is applied by the AFM tip, inducing local polarization switching. The simultaneously measured high-frequency electromechanical response (PFM spectroscopy) or subsequent imaging provides information of the formed domains. This approach allows nucleation to be explored at a predefined sample location. Alternatively, a homogeneous electric field applied via the top electrode of ferroelectric capacitors allows imaging of the statics and dynamics of the domain structure. In this case, the nucleation occurs at preferential defect sites and can be tuned by the amplitude of the applied field.[9] The time resolution of the experimental measurement on the domain radius development and the fraction of domain switching area has been enhanced with the development of high speed PFM.[10,11] Further investigation focused on single domain switching using a tip-generated inhomogeneous electric field.[12] The switching spectroscopy PFM detects the position of defects such as grain boundaries and allows the effect of the defect on switching mechanism to be observed. However, the limited spatial resolution (10~30 nm), and lack of structural information renders it virtually impossible to tie polarization dynamics to specific structural elements. Furthermore, whereas the intersection of the domain with the sample surface is visualized, only limited information is available on the depth profile of the switched region (i.e. extent of the domain in the direction normal to the film surface). Consequently, the nucleus volume and other parameters relevant to establishing the thermodynamic description of the switching process are unavailable.

In comparison, (scanning) transmission electron microscopy, (S)TEM, is a powerful tool which provides micro or even atomic-scale information on structural defects in oxide materials.[13] In ferroelectrics, TEM has been used to trace domain structure evolution[14,15] and phase transition[16,17] by imaging and electron diffraction during *in-situ* heating or cooling. Tan et al. designed an *in-situ* electric-field TEM holder to apply the bias to the whole sample and observed the general structural change, domain growth,[18] domain nucleation at grain boundaries in a polycrystal,[19] and electric field-induced fracture in plane-view direction.[20] These studies are complementary to the capacitor-based PFM measurements, and similarly the localization of nucleation cites cannot be independently controlled.

Here, we report direct observations of local structural changes through the thickness of ferroelectric film induced by bias using in-situ STEM. The concept is visualized in Fig. 1 showing the focused electron beam scanning the sample biased by a local probe. The combination of tip-electrode SPM and STEM allows visualization of domain growth through the thickness of the film during the switching, hence, the effect of pre-existing domain walls on nucleation and evolution of the new domain can be directly investigated. When combined with phase-field modeling the corresponding mesoscopic mechanisms can be deciphered.

## II. Domain structure in pristine material

As a model ferroelectric system, we selected 300 nm thick multiferroic $BiFeO_3$ (BFO) films epitaxially grown on (001) $DyScO_3$ (DSO) substrates (annealed at 1,200℃ for 3 hrs in flowing $O_2$)[21] by metal organic chemical vapor deposition (MOCVD). In the pristine state, the as-deposited films exhibit ordered arrays of 71° domain walls, corresponding to alternating in-plane domain orientations. The out of plane polarization component is uniform throughout the film. Fig. 2 shows surface topography and in-plane PFM images, consistent with the typical stripe ferroelectric domain pattern. The out-of-plane PFM image, Fig. 2(c), exhibits uniform contrast due to a preferred downward out-of-plane orientation.

*In-situ* measurement of electrical properties and observation at the nanometer level was carried out using a Nanofactory scanning probe microscope (SPM) holder as described in the Methods section. The electric field is localized below the tip and hence polarization switching is induced in a predefined volume, e.g. in the defect-free region or in the vicinity of a domain wall. Domain evolution is observed while the tip potential is increased with a step size of 100 mV. In order to optimize the contrast of the domain walls in both bright field (BF) and annular dark field (ADF) STEM images, the BF detector semiangle was adjusted to 5.56 mrad. In this imaging mode, the contrast in an epitaxial film of reasonably uniform thickness will be dominated by strain.[22] *Ex situ* studies of these samples showed no observable effect of electron beam related charging.

Fig. 3(a) shows the cross-section domain structure of a BFO thin film. The angle between the domain wall and the film/substrate interface is ~45º. This 71˚ domain wall separates two domains where one of the in-plane polarization components has the opposite sign, as illustrated in Fig. 3(a). The period of the ferroelastic domain pattern observed by STEM is 300 ~ 600 nm and is consistent with the PFM images in Fig. 2. 71º domain walls lie on {110} planes[23], so they can in principle be both normal and slanted with respect to the (001) surface of the substrate. However, we predominantly observe slanted walls as seen in Fig. 3(a), while normal walls are observed very rarely (see Supplementary Materials, Fig. S2).

The estimate for the expected width of the domain wall image using simple geometry (observation direction ~15˚ away from the [110]) and specimen thickness (100 nm) measured by Electron Energy Loss Spectroscopy is 50 nm. However, the observed width of the domain walls is somewhat smaller ($W_{domain\ wall}$ ~ 34 nm). This could happen because the domain walls are not exactly aligned with the (110) plane, possibly due to high depolarization field effects caused by close proximity of the top and bottom surfaces of a thin TEM sample.

### III. Nucleation and growth of a single domain

BFO has a rhombohedrally distorted perovskite structure with ferroelectric polarization oriented along one of the 8 pseudocubic [111] directions. Polarization switching by the field oriented along the [100] pseudocubic direction is thus a complex process, with a high degree of degeneracy between possible ferroelectric and ferroelastic pathways, selection between which will be affected by the presence of structural defects and pre-existing domain walls. To explore the interplay between the preexisting ferroelastic domain walls and polarization switching, we perform the switching using our localized STM tip at different positions with respect to domain walls.

Fig. 3(b-g) show the nucleation of a single domain, as manifested by a strain contour produced by the newly created bent ferroelastic domain wall, and its evolution with applied bias (images are selected from a set of 61; please see Supplemental movie 1 for the full set). Fig. 3(b) illustrates the geometry of the sample and the W tip before applying the bias. The tip apex was placed at the domain wall. At -0.8 V a domain nucleates and steadily increases in size under the applied bias. The threshold voltage for the domain nucleation varied in our experiments and was found to depend highly on the sample thickness. The shape of the newly nucleated domain is that of an oblate spheroid, as opposed to the classical needle-like domain shape expected for a depolarization field-driven process. The domain grows preferentially to the right side (Fig. 3(d)), indicating interference/pinning by the domain wall on the left. However, as the new domain approaches the right hand ferroelectric domain wall, the pinning effect is considerably smaller (Fig. 3(e)). The continuous increase of tip bias eventually results in the penetration of the new domain through the domain wall on the right, while the left hand domain wall continues to block growth (Fig. 3(f)). This sequence of the nucleation and growth events is illustrated schematically in Fig. 3(h). Note that the newly developed domain contrast disappears when the electric contact is removed (Fig. 3(g)), indicating that this switching is metastable. This is an intriguing result when considered in the context of PFM studies of ferroelectric ceramics and thin films., which find that without a conductive bottom electrode, such as in cases of bulk ceramics and single crystals of ferroelectrics, domain writing requires considerably larger bias values to accomplish,[24-27] in some cases, for thin films in particular, precipitating dielectric breakdown before such a bias is achieved.

However, in PFM an image of switched domains can only be obtained after removing the bias, while the *in situ* STEM approach allows us to do bias application and domain imaging simultaneously, revealing dynamic phenomena. We can thus suggest that the switching with and without bottom electrode proceeds similarly at similar bias values, while the stabilization of the new domains is greatly facilitated when the bottom electrode is present. It is also noteworthy that the metastable switching appears to be very slow, with domain structures developing over several hours. Thus it is possible to observe the gradual progression of the nucleation and growth events which suggests mechanisms of domain behavior. It is also notable that the original 71° domain walls remain visible during the entire process, indicating that one of the out-of-plane polarization components remains pinned throughout the experiment. When a bottom electrode is present, the tip-induced changes are rapid and persistent, and domain walls are mobile, as will be shown in a subsequent publication.[28]

The creation of the switched domain requires that the new polarization bound charge is compensated, however, for the case of nm-scale domain in a thin TEM sample the total charge required is miniscule (fractions of pC), amounting to ~fA current over the time scale of the experiment. This current can arise from a variety of mechanisms, from the residual leakage current through the BFO film to the surface conduction via carbon-rich top amorphous layer of the TEM sample created after ion milling and exposure to ambient contaminants.

The observed shape of the new domain can be explained using phase field modeling as shown in Fig. 4. Note that switching of the out-of-plane component results in a needle-like domain with a narrow straight ferroelectric wall (Fig. 4(b)). Since these walls are not associated with long range strain fields, as shown in Fig. 4(d), they are invisible in the STEM image. At the same time, the switching of the in-plane component of polarization $P_y$ results in a near–surface domain with a high-energy curved ferroelastic wall (Fig. 4(c)). This shape can be readily explained from the consideration of electrostatic depolarization fields that results in the elongation along the [100] or [010] axis, i.e. in the direction of the switched polarization component. Furthermore, recent first principles studies[29] of domain wall energies in BFO suggest that the domain wall energy is significantly lower for ferroelastic variants, facilitating formation of ferroelastic twins as opposed to ferroelectric domains. Thus in STEM images (Fig. 3) we are likely looking at the strain contour caused by the bent ferroelastic domain wall; this contrast pattern is substantially different from contrast produced by mechanical contact (indentation) of the tip with no applied bias (see Fig. S3 in Supplementary Materials).

However, in the cross-section geometry of our sample the depolarization field conditions are substantially different from a planar film (confined not only by film thickness, but also by specimen thickness) and need to be taken into account explicitly. Therefore, to explain the interactions of the growing domains with the existing 71˚ domain walls, a new set of phase field simulations was performed, for the "nanowire" geometry that more accurately describes a TEM sample. Fig. 5(a) shows the simulated 3-dimensional geometry of the domain structure after switching. A tip with voltage of -20 V was applied at the center of the domain. The nucleation and growth of a ferroelectric domain at the 'a' side of the domain wall observed in the experiment was reproduced in the cross section plane of the [110] oriented nanowire (used as a model for the TEM sample) (Fig. 5(b)). As the domain grows, the domain wall tilts along the 71° domain wall (Fig. 5 (c)), i.e. the 71° domain wall inhibits the domain growth, similar to the experimental observation (Fig. 3).

## IV. Nucleation and growth of multiple domains

When the contact area between the W tip and the sample surface is comparable with domain size, multiple domains form nearby. However, many of the same trends found in the earlier experiment apply. When the tip makes a mechanical and electrical contact without a bias, the preexisting 71° domain walls are clearly visible, Fig. 6(a). Since the sample thickness (along the electron beam direction in TEM) of the observed area is much larger (230 nm), the first nucleation event occurred at a higher bias of 6.5 V (Fig. 6(b)). The first domain forms at an acute angle intersection between the domain wall and the surface as highlighted in green in Fig. 6(f). As the experiment progresses, the second domain forms at the next acute angled site of the neighboring domain II as shown in Fig. 6(c, g). Finally, a third domain appears at the obtuse angle region of domain I, directly under the tip (Fig. 6(d, h)). This domain (red) coalesces with nearby domain (green) which is expanding from the left side in the end, as shown in Fig. 6(e, i).

The preferential nucleation site at the acute angle between the domain wall and the surface (site 'a' in Fig. 3(h)) to the opposite side making an obtuse angle (site 'o') is consistent with the single domain development in Fig. 3; it was also observed in repeated experiments. Thus the potential for the nucleation appears to depend on the position with respect to the domain wall, with the global minimum at the acute angle site, and a smaller local minimum at the obtuse angle side, which is consistent with the theoretical calculation showing the asymmetric potential dip for the domain nucleation near the twin domain boundary[30]. Hence, one direction is preferred for charged wall motion, resulting in intriguing asymmetric polarization dynamics.

This behavior also manifests as a directionality of the domain wall pinning effect. The growth of the new domain across the domain wall is rather easier when the other side of the wall is a preferential nucleation site (is at an acute angle with the surface). In contrast, the pinning effect is stronger when the other side of the wall is a non-preferred nucleation site. Depending on the tip shape and position, a new domain can nucleate at the non-preferred obtuse angle site before the first domain can penetrate the domain wall as shown in Fig. 6(d). This indicates the activation energy for passing through the 71° domain in a non-preferred direction can be higher than the nucleation barrier for a new domain.

## VI. Summary

To summarize, *in-situ* nucleation and growth of ferroelastic domains was directly visualized using strain contrast in STEM bright field images. The slow kinetics in metastable switching of BFO/DSO material allowed us to unveil previously unknown domain switching mechanisms, including preferential nucleation sites and asymmetric pinning at ferroelastic 71° domain walls, factors that are highly relevant to multiaxial ferroelectric and multiferroic materials. The tip-applied local bias allows polarization switching to be induced at a predefined location, and the mechanisms of domain wall growth and domain-defect interactions can therefore be explored. The observed asymmetric domain – domain wall interactions suggest the intriguing possibility of rectifying polarization-based logic devices incorporating such a diode-like function.

Beyond ferroelectric and multiferroic materials, this experimental setup can serve to induce ionic motion, vacancy injection and subsequent ordering in materials with mobile cations or vacancies.[31] The addition of high-resolution e-beam crystallography, electron energy-spectroscopy based chemical imaging,[2] and ultimately atomically-resolution structural imaging capabilities[32, 33] will allow bias-induced dynamic processes to be explored at the mesoscopic, single-defect and ultimately atomic levels. This will provide much-needed information to visualize, understand, and consequently optimize functionality of energy storage and generation devices through nanoscale control of electrochemical defect functionality.

## Acknowledgements

This research is sponsored by the Materials Sciences and Engineering Division, Office of Basic Energy Sciences of the U.S. Department of Energy, and by appointment (H.J.C.) to the ORNL Postdoctoral Research Program administered jointly by ORNL and ORISE. Instrument access via Oak Ridge National Laboratory's SHaRE User Facility, which is sponsored by the Scientific User Facilities Division, Office of Basic Energy Sciences, the U.S. Department of Energy, is gratefully acknowledged. The work at Berkeley is also partially supported by the Semiconductor Research Corporation–Nanoelectronics Research Initiative–Western Institute of Nanoelectrics program. The work at Penn State is supported by DOE Basic Sciences under grant number DE-FG02-07ER46417.

**Figure Captions**

Figure 1. Artistic vision of the concepts in this study. The confinement of an electric field by an SPM probe allows a bias-induced phase transition to be probed. STEM imaging allows observations of the probe and dynamic changes in the structure of the sample.

Figure 2. (a) AFM surface topography (b) in-plane and (c) out-of-plane PFM image of the domain structure of the as-grown BFO/DSO heterostructures.

Figure 3. (a) BF STEM cross-section image showing the 71° domain wall structure of a BFO film on DSO. (b-g) Consecutive images showing bias induced domain nucleation and growth. The arrows indicate 71° domain walls: (b) original domain structure before applying the bias, (c) domain nucleation, (d) asymmetric growth due to pinning by the pre-existing domain wall on the left, (e) penetration of the domain wall on the right, (f) the new domain reaches the substrate and growth saturates, (g) original domain structure is recovered after the bias is removed. (h) Schematics representing the sequence of the domain nucleation and growth events in (b-f).

Figure 4. Phase field modeling of the switching in BFO thin films. (a) Top view of the orthogonal switched domains with out-of-plane (pink-$P_z$) and two in-plane (green-$P_y$ and blue-$P_x$) orientations, note partial overlap. (b-c) Side view of the BFO film showing the out-of-plane component ($P_z$) which is normal to the tip (b), in-plane component ($P_y$) (c) and elastic energy (d). The new domain walls induced by out-of-plane switching are not associated with a long range strain field, while in-plane polarization switching produces hemispherical domains with a strong strain signature.

Figure 5. (a) A 3-dimensional geometry of the multi-domain structure for phase field simulation. (b) The sequential cross section plots of the [110] oriented BFO "nanowire", showing the nucleation and growth of the bias induced domain in the thin TEM sample. (c) Vector plot showing polarization rotation due to pinning by the pre-existing domain wall on the left.

Figure 6. (a-e) BF STEM images showing the nucleation and growth of multiple domains. (f-i) corresponding colorized images of (b-e), respectively.

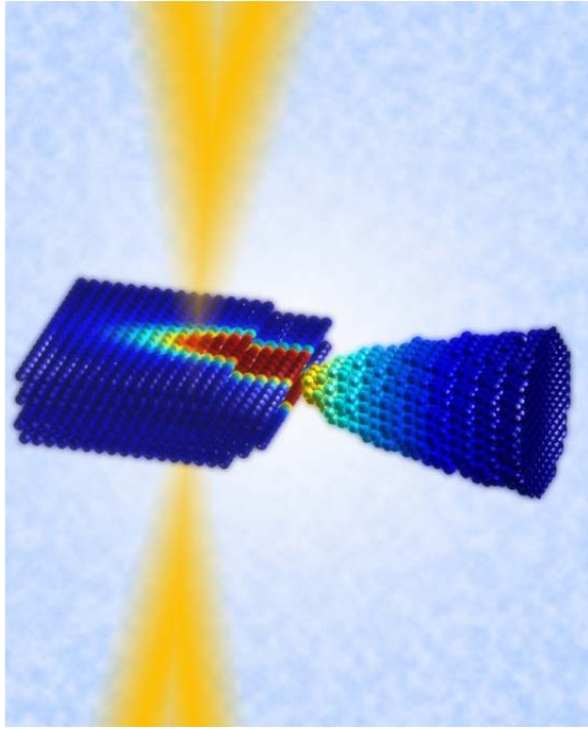

Fig. 1

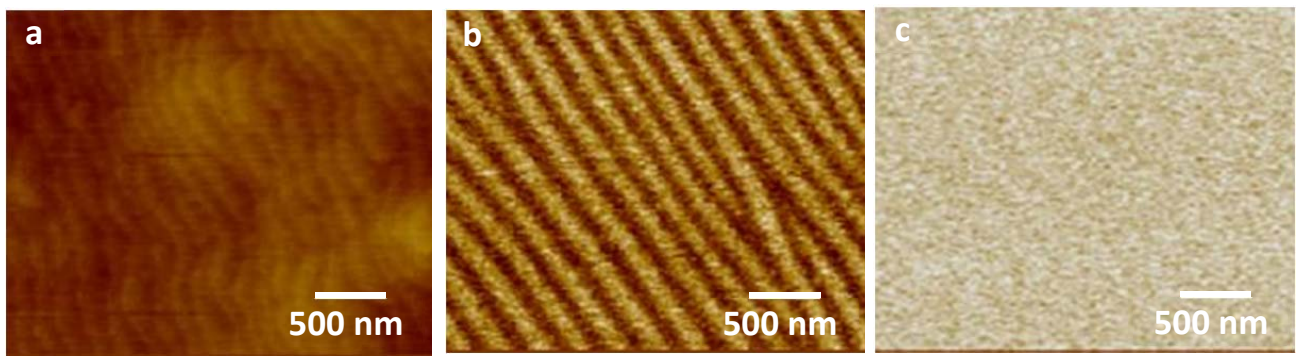

Fig. 2

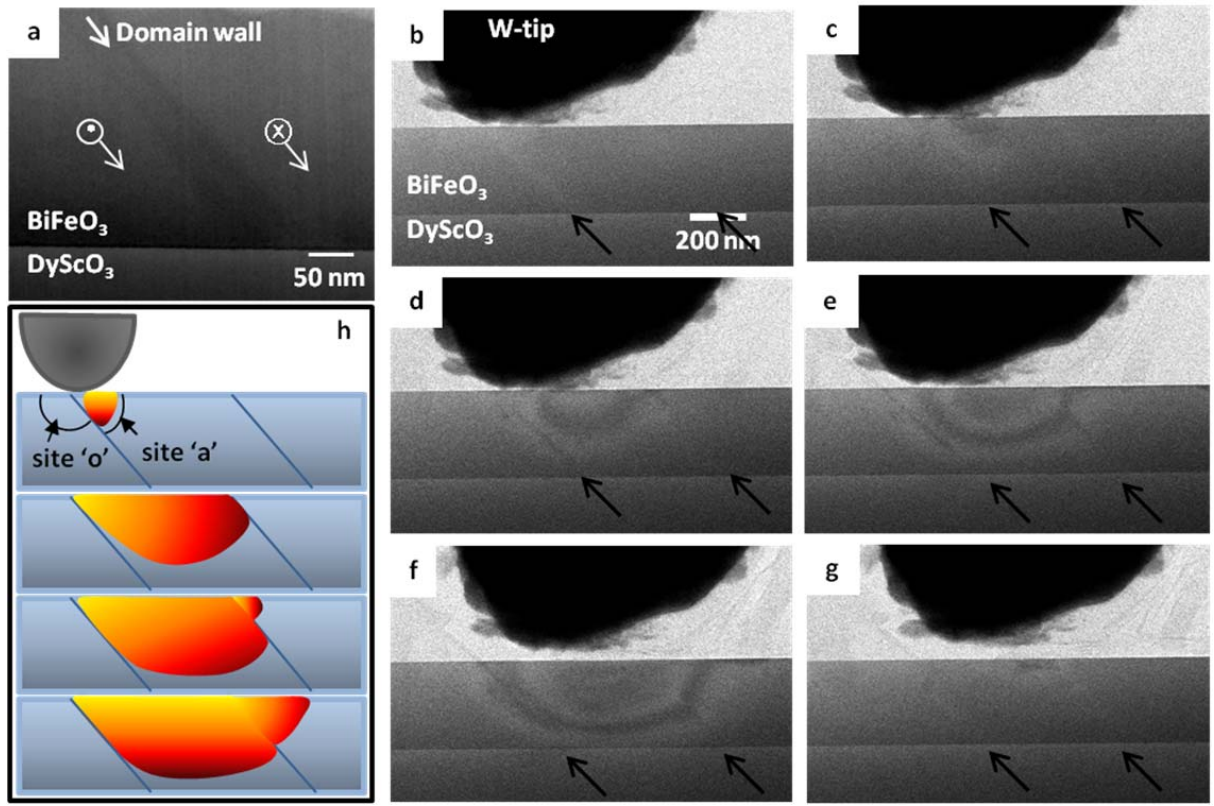

Fig. 3

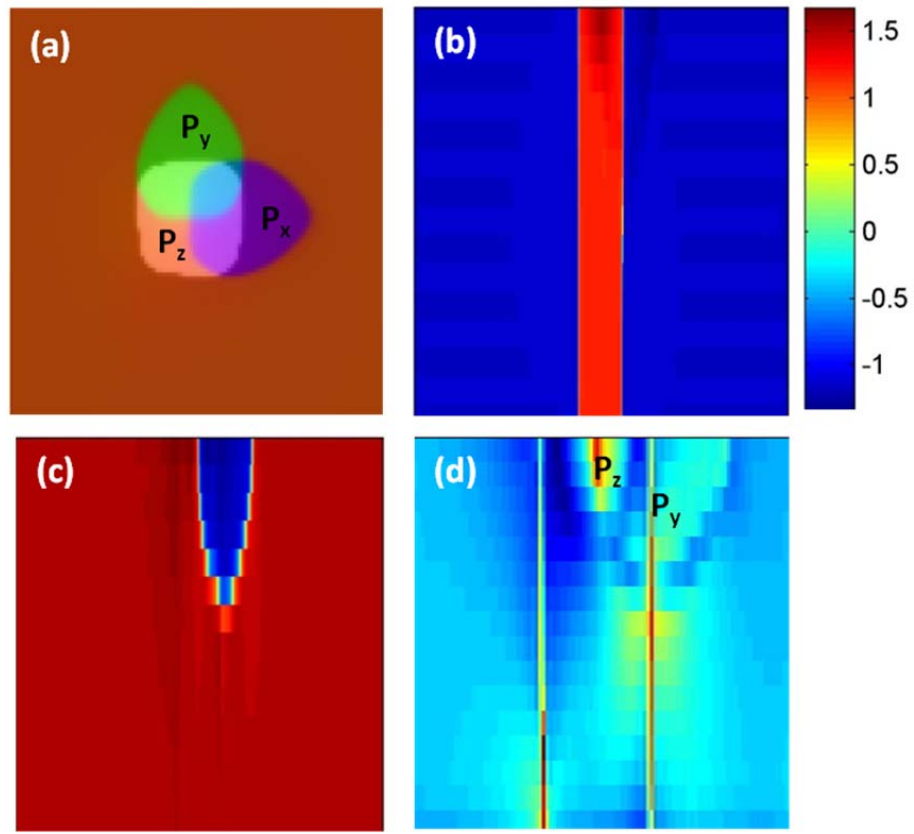

Fig. 4

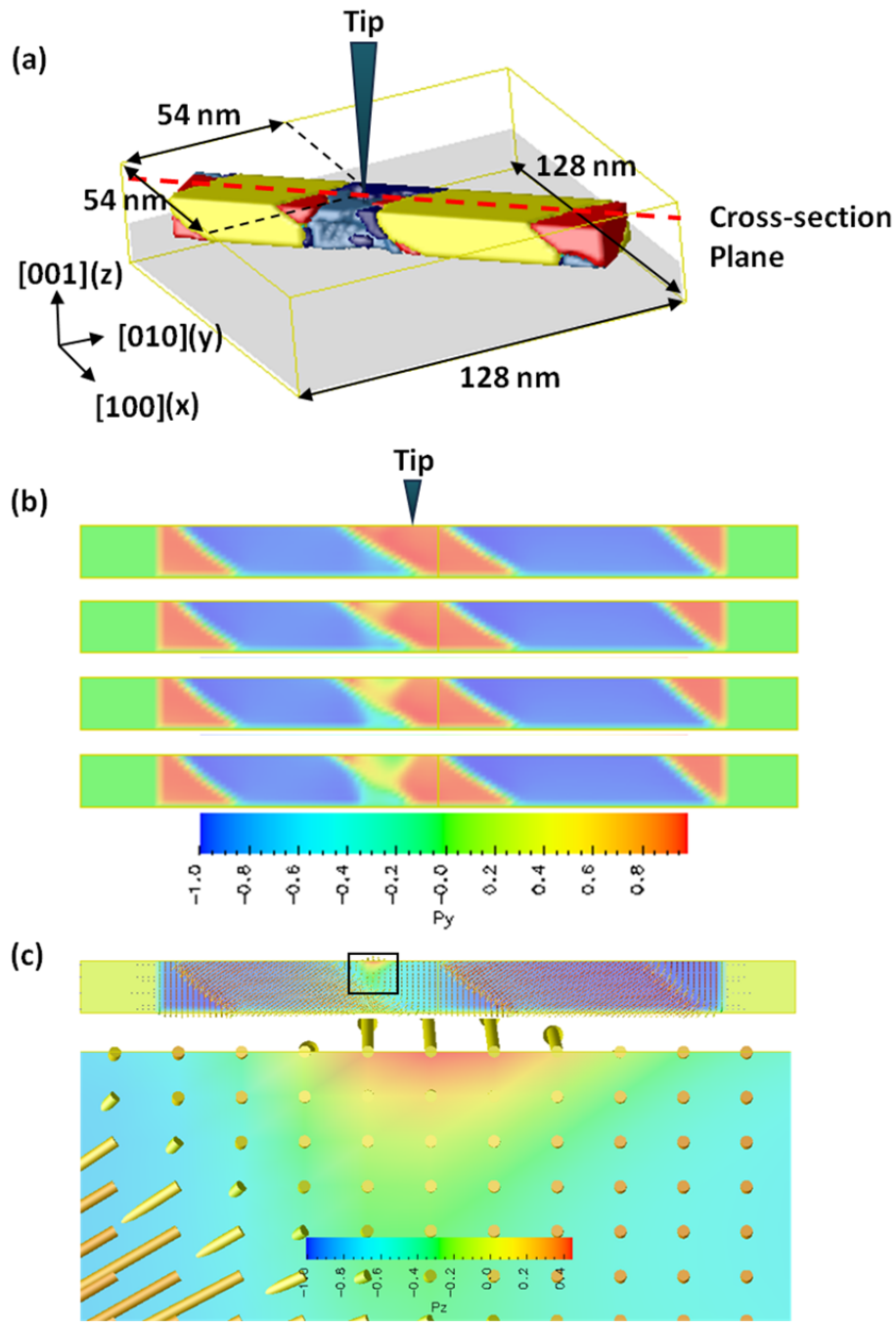

Fig. 5

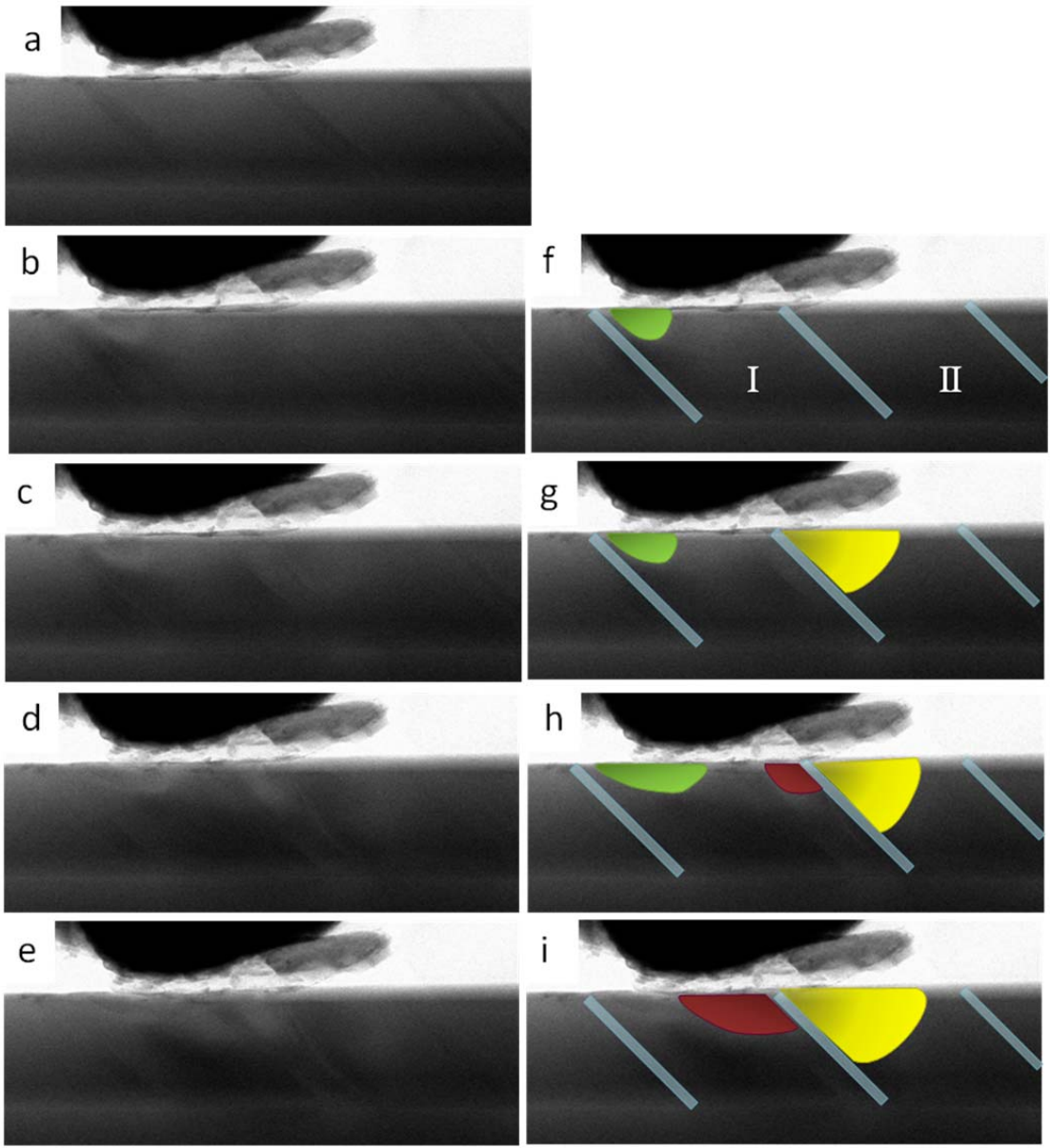

Fig. 6